\documentclass[onecolumn,noshowpacs,nofootinbib,colorlinks,hyperindex,unicode,10pt]{revtex4-2}
\usepackage{graphicx}
\usepackage{tikz}

\usepackage{simpler-wick}

\usepackage[compat=1.1.0]{tikz-feynman}
\usepackage{verbatim}
\usepackage{bbm}
\usepackage{amsfonts,amsmath,amssymb,tikz,accents}
\usepackage[eulergreek]{sansmath}
\usepackage{indentfirst,bigints}
\usepackage{hyperref,upgreek}
\usepackage{bm,graphics,color}
\usepackage{feynmp-auto}
\usepackage{slashed}
\counterwithout{equation}{section}
\usepackage{hyperref}
\usepackage{pgfplots}
\pgfplotsset{compat=1.16}

\hypersetup{hidelinks,backref=true,pagebackref=true,hyperindex=true,colorlinks=true,breaklinks=true,urlcolor= blue}
\hypersetup{%
  colorlinks = true,
  linkcolor  = blue,
  citecolor = cyan,
}
\usepackage{textgreek}
\usepackage{color,xcolor}
\newcommand{\gdualn}[1]{\overset{\:{}^{{}^{\boldsymbol{\neg}}}}{\smash[t]{#1}}} 

\def\0{\mbox{\boldmath$\displaystyle\mathbb{O}$}}
\usepackage{bbold}
\def\I{\openone}
\def\openone{\mathbb I}

\def\p{\mbox{\boldmath$\displaystyle\boldsymbol{p}$}}

\newcommand{\orcidicon}{%
	\begin{tikzpicture}
	\draw[lime, fill=lime] (0,0)
		circle [radius=0.16]
		node[white] {{\fontfamily{qag}\selectfont \tiny ID}};
	\draw[white, fill=white] (-0.0625,0.095)
		circle [radius=0.007];
	\end{tikzpicture}	\hspace{-2mm}
}
\newcommand\orcidg{{\href{https://orcid.org/0000-0002-7942-7941}{\orcidicon}}}
\newcommand\orcidRR{{\href{https://orcid.org/0000-0002-8283-2577}{\orcidicon}}}
\newcommand\orcidLF{{\href{https://orcid.org/0000-0002-9186-2807}{\orcidicon}}}

\usepackage{float,xcolor,upgreek}
\usepackage{color} 
\usepackage{tikz-feynman}
\tikzfeynmanset{compat=1.0.0}
\newcommand{\beq}{\begin{eqnarray}}
\newcommand{\eeq}{\end{eqnarray}}
\newcommand{\bea}{\begin{eqnarray}}
\newcommand{\eea}{\end{eqnarray}}

\begin{document}

\title{Unraveling the Physical Meaning Behind Elko's Dual structure}

\author{Gabriel Brandão de Gracia\orcidg{}}
\affiliation{Departamento de Física, Universidade Federal do Triângulo Mineiro UFTM,\\
38.025-180, Uberaba, MG, Brazil}
\email{gabriel.gracia@uftm.edu.br}

\author{Rodolfo José Bueno Rogerio\orcidRR{}}
\affiliation{Centro Universitário UNIFAAT, Atibaia-SP, 12954-070, Brazil.}
\email{rodolforogerio@gmail.com}

\author{Luca Fabbri\orcidLF{}}
\affiliation{DIME, Sez. Metodi e Modelli Matematici, Universit\`{a} di Genova,\\
Via all'Opera Pia 15, 16145 Genova, Italy. \\ GNFM, Istituto Nazionale di Alta Matematica, P.le Aldo Moro 5, 00185 Roma, Italy}
\email{fabbri@dime.unige.it}

\begin{abstract}
\indent In this work, we shall analyze the necessity of a proper definition of the dual structure for Elko spinors, and singular spinors in general. We examine in detail why the Dirac dual structure fails and it is not functional for these cases, highlighting all the physical consequences of this misdirection.
The approach considered here is different from the one usually taken in current literature. We pinpoint the shortcomings that this dual structure brings to particle interpretation, and propagator structure, as well as the implications in non-locality, and the existence of negative energy levels. This investigation furnishes a background for the successful achievements in the field associated with such proper dual formulation.
\end{abstract}

\maketitle


\section{Opening Section}

The Elko\footnote{Elko is a German acronym for \emph{Eigenspinoren des Ladungskonjugationsoperators}} spinors were mathematically proposed in the early 2000s \cite{jcap, dharamprd}. Such spinors define a complete set of eigenspinors of the charge conjugation operator. Due to their inherent characteristic, they are strong candidates for describing dark matter \cite{jcap}. Since their discovery, these mathematical objects have been extensively explored in several contexts such as in mathematical physics \cite{polarform2020,elkopolar,chenggeneral,chenglagrangian, juliodoublets}, cosmology \cite{saulo1,saulo2,saulo3, cylcosmelkology}, phenomenology \cite{roldao2023,roldaofermionic,julioperturbative,chengyukawa}, and other related areas; implying new and deeper insights in theoretical dark matter investigations. By bringing forth an algebraic structure different than the standard Dirac one, they have compelled the scientific community to revisit the physical and mathematical foundations defining the spinor representations of the Poincaré group, in order to achieve a correct classification of the associated particles in compliance with the previous seminal works of Wigner \cite{wigner1, wigner2,  dharamnpb}, Weinberg \cite{weinberg1}, and Ryder \cite{ryder}.

The renowned theory of Dirac spinors has been well-established since its conception around 1928 \cite{pamdirac}. However, as we have just mentioned, the structure of Dirac spinors appears not to be unique \cite{rrdirac1, rrdirac2}. Regarding Dirac spinors, they are based on the link between the representation spaces $(0, 1/2)\oplus(1/2,0)$, provided by the parity operation, as it can be seen in details in \cite[Chapter 2]{ryder}, proved by Weinberg as the well-known no-go theorem \cite{weinberg1}, and recently re-examined in \cite{dharamnogo, rodolfonogo}. Then, one may pose the following question: Is this the only way to correlate these representation spaces? The answer to such a question is no. This was demonstrated by Ahluwalia, when he formally introduced the Elko spinors in \cite{jcap} and also in other correlated papers \cite{dharamnpb, dharambosons, dharamspinstatistic}. To achieve this, Ahluwalia abstains from parity symmetry and, instead, considers an algebraic relation among the Pauli ($\sigma_x$, $\sigma_y$ and $\sigma_z$) matrices, commonly referred to as the ``\emph{Magic of Pauli matrices}'' \cite{ramond}. Since the link between representation spaces is not unique, one can raise another question: What physical consequences can this bring about? Well, as can be observed in the extensive literature on Elko spinors \cite{rodolfoconstraints, rjdual, rrtaka, propagatormpla,propagatorbsm, rjtau, chengtipo4, out, jrold1, rog1, rog2, beyondlounesto}, one must revisit the study of dual prescription to define a structure capable to provide relevant physical information for the theory. By physical information we mean non-vanishing and real Lorentz invariant orthonormal relations, invariant spin-sums, and local associated quantum fields.   

As it is well known, the Dirac spinors stand for eigenspinors of the parity operator \cite{aaca}. Accordingly to Speran\c{c}a in \cite{speranca}, it is possible to define an operator that acts on the functional form as well as on the spinor argument, the Dirac operator generates the covariant parity operator (this being a mathematical identity, namely $\mathcal{P}=m^{-1}\gamma_{\mu}p^{\mu}$, meaning that $\mathcal{P}\psi = \pm \psi$. Thus, when defining the dual structure, Dirac considered the following $\bar{\psi}=\psi^{\dag}\gamma_0$, responsible for establishing a Lorentz invariant quantity $\bar{\psi}\psi$. However, is the Dirac dual structure a universal one that can be used for any spinor? We must emphasize that the correct answer is no! It is not universal and should not be used for all spinor classes defined by Lounesto classification \cite{lounestolivro}. The reason for this will become clear as our discussion progresses.

Naively, by employing such a dual structure for all kinds of spinors, one might lose information about the theory in question \cite{aaca, mdobook}. We should clarify the following: since Dirac spinors are eigenstates of the parity operator, we could then rewrite the dual structure for the free field as
\begin{eqnarray}\label{dualgeral}
\bar{\psi}=[\mathcal{P}\psi]^{\dag}\gamma_0,
\end{eqnarray}
and since $\mathcal{P}\psi=\pm\psi$, the above relation yields $\bar{\psi}=\psi^{\dag}\gamma_0$. Thus, one can prove that the most general manner to write the Dirac dual structure is given by \eqref{dualgeral}. 

With that being said, we must emphasize that such observations have already been scrutinized in \cite{jcap}, where the author seeks a redefinition of the dual in order to establish a suitable structure encoding the correct physical properties. In this manner, from 2004 until today, the research on Elko spinors is based on a redefined dual formulation \cite{rrdirac1, rrdirac2, rjdual, rrtaka, jrold1, rog1, rog2}, ensuring meaningful results such as locality of the associated quantum field, Hermiticity \cite{elkoherm}, the correct particle interpretation, Takahashi inversion theorem \cite{rrtaka}, among other important aspects within general relativity framework, for Elko higher spin theories, gravitational interaction and also supergravity formulation\footnote{The connection with supergravity can be established by introducing an antisymmetric field playing the role of the square root of the Riemann tensor.   }  \cite{supergrav} besides other applications \cite{nieto1, nieto2} .\\
\indent Having established such a formulation, one can also address this discussion by a different approach.
In addition to the success brought by the correct definition of the dual structure given in \eqref{dualgeral}, one can proceed in the opposite direction and ask: which are the physical implications of considering the standard dual definition for the theory associated with fields constructed upon the Elko spinor coefficients? This is the goal of the present investigation. We will show that it leads to a problematic situation in which the quantum field propagator obtained by the vacuum average of the field's two-point time-ordered product is not proportional to the inverse of the Lagrangian differential operator. Moreover, we also show that a possible redefinition of such an operator to ensure a correspondence between these objects necessarily leads to a non-local model, violating an important quantum field theory (QFT) foundation.\\
\indent Finally, we consider the new basis for the quantum field, associated with the Wigner degeneracy due to the anti-linear nature of the charge conjugation operator, ensuring full compliance with Weinberg conditions for a particle representation \cite{weinberg1,wigner1, dharamnpb, elkostates}, to demonstrate that the use of the standard Dirac dual implies in Hamiltonian instabilities. More specifically, the Hamiltonian operator is Hermitian but possesses eigenstates with negative energy. Beyond this point, the one-particle states associated with the quantum field expansion are not Hamiltonian eigenstates anymore, violating the standard particle interpretation of QFT.\\
\indent Although the Elko spinor norms vanish under the Dirac dual, the mass term constructed from the quantum fields is different from zero. However, we highlight that it leads to the set of problematic features described above. We also investigate the compliance of such an approach with the correspondence principle and derive considerations on wider possibilities for the operator algebra. Then, our objective is to provide sharper definitions of such discussions and highlight shortcomings associated with the Dirac adjoint prescription to reinforce the reasons for the historical path followed by the pioneers of Elko spinor investigations.\\
\indent The paper is organized as follows: in the Sec. \ref{set} , we introduce the basis associated with the rotational invariance, the definition of singular spinors as well as the structure of the quantum field operator. In Sec. \ref{2}, the spin sums, the spinor products, and the Lagrangian for the case of the Dirac adjoint is derived for the whole set of singular spinors. In Sec. \ref{3} we show that this prescription for the dual leads to a problematic mismatch between different approaches when we define the propagator, as it is possible to see, such a quantity evinces a non-local feature. We also make some discussions on the Hamiltonian operator. We explicitly obtain the structure of such an operator and provide several discussions under the light of the standard particle interpretation in quantum field theory. In the section \ref{4}, we conclude our formalism.

\section{Spinorial set-up} \label{set}
We start this section by defining the central aspects of the general singular spinors. They are displayed below \footnote{The lower indexes, $_h$ and $_h^{\prime}$, stands for the helicity of right-hand and left-hand components, respectively. Such a prescription carries pedagogical and historical purposes. Nonetheless, we call attention to the following fact, in the recent literature, \emph{e.g.} \cite{mdobook,jhepgabriel}, authors make the following replacement: $\uplambda^S_{-} \rightarrow\uplambda^S_{\{+,-\}}$, $\uplambda^S_{+}\rightarrow\uplambda^S_{\{-,+\}}$, $\uplambda^A_{+}\rightarrow\uplambda^A_{\{+,-\}}$ and $\uplambda^A_{-}\rightarrow\uplambda^A_{\{-,+\}}$.  }
\begin{subequations}
\begin{eqnarray}\label{espi}
\uplambda^S_{\{+,-\}}(\boldsymbol{0})=\left(\begin{array}{c}
\alpha\Theta\upphi^{-*}(\boldsymbol{0}) \\ 
\beta\upphi^{-}(\boldsymbol{0})
\end{array} \right),
\qquad\quad \uplambda^S_{\{-,+\}}(\boldsymbol{0})=\left(\begin{array}{c}
\alpha\Theta\upphi^{+*}(\boldsymbol{0}) \\ 
\beta\upphi^{+}(\boldsymbol{0})
\end{array} \right),
\\\label{nor} 
\uplambda^A_{\{+,-\}}(\boldsymbol{0})=-\left(\begin{array}{c}
-\alpha\Theta\upphi^{+*}(\boldsymbol{0}) \\ 
\beta\upphi^{+}(\boldsymbol{0})
\end{array} \right),
\qquad\quad \uplambda^A_{\{-,+\}}(\boldsymbol{0})=\left(\begin{array}{c}
-\alpha\Theta\upphi^{-*}(\boldsymbol{0}) \\ 
\beta\upphi^{-}(\boldsymbol{0})
\end{array} \right),
\end{eqnarray}
\end{subequations}
where the upper indexes $S$ and $A$ refer to the fact that when reaching the Elko configuration with $\alpha=i$ and $\beta=1$, the associated spinors become self-conjugate and anti-self-conjugate with relation to the action of the charge conjugation operator, respectively. The spinors introduced above carry a vanishing norm under the Dirac adjoint prescription, accordingly observations in \cite{lounestolivro}.
The spinorial components are given by
\begin{eqnarray}\label{components}
\upphi^{+}(\boldsymbol{0}) = \sqrt{m/2}\left(\begin{array}{c}
1 \\ 
0
\end{array}\right), \qquad\qquad\qquad  \upphi^{-}(\boldsymbol{0}) = \sqrt{m/2}\left(\begin{array}{c}
0 \\ 
1
\end{array}\right),
\end{eqnarray}
and $\Theta$ stands for the Wigner time-reversal operator \cite{ramond}, which in the spin $1/2$ representation, yields 
\begin{eqnarray}
\Theta = \left(\begin{array}{cc}
0 & -1 \\ 
1 & 0
\end{array}\right).
\end{eqnarray} 
 
As recently observed, singular spinors are endowed with Wigner degeneracy, thus, we also introduce the following set of degenerated spinors, namely $\rho$ spinors, which reads  
\begin{subequations}
\begin{eqnarray}\label{espi1}
\rho^S_{\{+,-\}}(\textbf{0})&=&\left(\begin{array}{c}
\beta^*\upphi^{+}(\boldsymbol{0}) \\ 
\alpha^*\Theta\upphi^{+*}(\boldsymbol{0})
\end{array} \right),\qquad\quad \rho^S_{\{-,+\}}(\textbf{0})=\left(\begin{array}{c}
\beta^*\upphi^{-}(\boldsymbol{0}) \\ 
\alpha^*\Theta\upphi^{-*}(\boldsymbol{0})
\end{array} \right),\\
\rho^A_{\{+,-\}}(\textbf{0})&=&\left(\begin{array}{c}
-\beta^*\upphi^{-}(\boldsymbol{0}) \\ 
\alpha^*\Theta\upphi^{-*}(\boldsymbol{0})
\end{array} \right),
\qquad\quad \rho^A_{\{-,+\}}(\textbf{0})=  -\left(\begin{array}{c}
-\beta^*\upphi^{+}(\boldsymbol{0}) \\ 
\alpha^*\Theta\upphi^{+*}(\boldsymbol{0})
\end{array} \right).\label{nor1}
\end{eqnarray}
\end{subequations}
after a boost, the spinors are written as $\lambda_h(\textbf{p})= \kappa\lambda_h(\textbf{0})$ and $\rho_h(\textbf{p})= \kappa\rho_h(\textbf{0})$ in an arbitrary frame. The boost operator for the $\left(\frac12,0\right)\oplus\left(0,\frac12\right)$ representation
\begin{eqnarray}
\kappa = \sqrt{\frac{E+m}{2m}}\left(\begin{array}{cc}
\I+ \frac{\vec{\sigma}\cdot\vec{\textbf{p}}}{E+m} & 0 \\ 
0 & \I- \frac{\vec{\sigma}\cdot\vec{\textbf{p}}}{E+m}
\end{array} \right).
\end{eqnarray} 

\indent An important characteristic of singular spinors in general, is the fact that they do not obey the Dirac-like equations, as previously verified in \cite{,jcap, rcampla}. For the specific case of Elko spinors, they obey a concise system of coupled relations, which reads \footnote{The parameter $m$ is identified as the mass of the Elko spinor.}
\bea \slashed{p}\uplambda^S_{\{\mp,\pm\}}(\boldsymbol{p})=\pm im \uplambda^S_{\{\pm,\mp\}}(\boldsymbol{p}), \qquad \qquad \slashed{p}\uplambda^A_{\{\pm,\mp\}}(\boldsymbol{p})=\mp im \uplambda^A_{\{\mp,\pm\}}(\boldsymbol{p}) \label{Elkop},\eea
\indent Interestingly enough, these relations imply that all Elko spinors are in the kernel of the Klein-Gordon differential operator, characterizing them as good candidates to define a particle description. It is worth mentioning that the on-shell nature holds for all singular spinors.

\indent In order to define a spinor basis compatible with the rotational constraint, that should be fulfilled to correctly describe one particle states \cite{weinberg1, dharamnpb, jhepgabriel, rrsingular2024}, one introduces the set $\xi_h(\boldsymbol{\p})$ related to the particle sector
 \begin{eqnarray}
&&\xi_1(\boldsymbol{\p}) = \uplambda^S_{\{+,-\}}(\boldsymbol{\p}),\label{wig1} \qquad\quad \xi_2(\boldsymbol{\p}) = \uplambda^S_{\{-,+\}}(\boldsymbol{\p}),\\
&&\xi_3(\boldsymbol{\p}) = \rho^S_{\{+,-\}}(\boldsymbol{\p}), \quad\qquad \xi_4(\boldsymbol{\p}) = \rho^S_{\{-,+\}}(\boldsymbol{\p}).
\end{eqnarray}
Similarly, the set  $\upchi_h(\boldsymbol{\p})$, related to the anti-particle sector, is defined below
\begin{eqnarray}
&&\upchi_1(\boldsymbol{\p}) = \uplambda^A_{\{+,-\}}(\boldsymbol{\p}), \quad\qquad \upchi_2(\boldsymbol{\p}) = \uplambda^A_{\{-,+\}}(\boldsymbol{\p}),\\
&&\upchi_3(\boldsymbol{\p}) = \rho^A_{\{+,-\}}(\boldsymbol{\p}), \qquad\quad \upchi_4(\boldsymbol{\p}) = \rho^A_{\{-,+\}}(\boldsymbol{\p}).
\label{wig2}\end{eqnarray}

Interestingly, when Elko spinors are considered, this construction separates the particle and antiparticle sets as self-conjugate or anti-self-conjugate with relation to the action of the charge conjugation operator, in accordance with the Wigner degeneracy \cite{dharamnpb, jhepgabriel, rrsingular2024}.\\
\indent Then, the quantum field expansion in terms of the singular spinor expansion coefficient functions,  read \cite{rrsingular2024}
\begin{eqnarray}\label{campoquanticofinal}
\uplambda(x) =\int\frac{d^3 p}{(2\pi)^3}
\frac{1}{\sqrt{ m(|\alpha|^2+|\beta|^2) E(\p)}}
\bigg[
\sum_{\textit{h}=1}^{4} {c}_{\textit{h}}(\p)\xi_{\textit{h}}(\p) e^{-i p_\mu x^\mu}+ \sum_{\textit{h}=1}^{4} d^\dagger_{\textit{h}}(\p)\upchi_{\textit{h}}(\p) e^{i p_\mu x^\mu}\bigg],
\end{eqnarray}
and the associated dual in terms of the Dirac adjoint is defined below \footnote{Considering $\bar{\xi}_{\textit{h}}(\textbf{p}) = \xi_{\textit{h}}(\textbf{p})^{\dagger}\gamma_0$ and $\bar{\upchi}_{\textit{h}}(\textbf{p}) = \upchi_{\textit{h}}(\textbf{p})^{\dagger}\gamma_0$}
\begin{eqnarray}\label{campoquanticodualfinal}
\bar{\uplambda}(x) = \int\frac{d^3 p}{(2\pi)^3}
\frac{1}{\sqrt{ m (|\alpha|^2+|\beta|^2) E(\p)}}
\bigg[
\sum_{\textit{h}=1}^{4} c^\dagger_{\textit{h}}(\p)\bar{\xi}_{\textit{h}}(\p) e^{i p_\mu x^\mu}+ \sum_{\textit{h}=1}^{4} d_{\textit{h}}(\p)\bar{\upchi}_{\textit{h}}(\p)e^{-i p_\mu x^\mu}\bigg],
\end{eqnarray} 
here, $p_\mu=(p_0,\p)$, with $p_0=\sqrt{\p^2+m^2}$ displaying the dispersion relation compatible with the Klein-Gordon equation. Interestingly, setting $\alpha=i$ and $\beta=1$, one reaches the well-known Elko's quantum field operators with the Dirac dual instead of its proper well-defined dual structure \cite{dharamnpb, jhepgabriel}.\\
 \indent The non-vanishing fermionic anti-commutative relations among the creation and annihilation operators are highlighted below
\begin{equation}
\Big\{c_{\textit{h}}(\p),c^{\dagger }_{\textit{h}^{\prime}}(\p^{\prime})\Big\} =(2\pi)^3 \delta^3(\p-\p^\prime)\delta_{\textit{h}\textit{h}^{\prime}}. \label{anticomutador}
\end{equation}
Note that each $c_{\textit{h}}(\p)$ and $c^{\dagger }_{\textit{h}}(\p)$ operator creates (or annihilates) an specific type of particle with label $h$ and momentum $\p$. It is worth mentioning that the relations above are also valid for the pair $d_{\textit{h}}(\p)$ and $d^{\dagger }_{\textit{h}}(\p)$, associated with the antiparticle sector.

\section{Properties of the Singular Spinor with Dirac Adjoint}\label{2}

This section is reserved for a deep understanding of the implications of the dual structure.  
For the purpose of the work, we set the Dirac adjoint structure as $\bar{\psi}_{\textit{h}}(\textbf{p}) = \psi_{\textit{h}}(\textbf{p})^{\dag}\gamma_0$, and verify the physical consequences. Thus, for the case at hand, the previously introduced basis acquires the following dual structures 
$\bar{\xi}_{\textit{h}}(\textbf{p}) = \xi_{\textit{h}}(\textbf{p})^{\dag}\gamma_0$ and $\bar{\upchi_{\textit{h}}}(\textbf{p}) = \upchi_{\textit{h}}(\textbf{p})^{\dag}\gamma_0$. This definition implies in vanishing norms for all the basis elements
\begin{eqnarray}
&&\bar{\xi}_{\textit{h}}(\p)\xi_{\textit{h}}(\p) = 0, \label{ortoF1}
\\
&&\bar{\upchi}_{\textit{h}}(\p)\upchi_{\textit{h}}(\p) = 0. \label{ortoF2}
\end{eqnarray}
whereas the label sums have the following structure
\begin{eqnarray}
&& \sum_{\textit{h}} \xi_{\textit{h}}(\p)\bar{\xi}_{\textit{h}}(\p)  = \frac{\big(|\alpha|^2+|\beta|^2\big)}{2}\gamma_{\mu}p^{\mu}, \label{spinsums1}
 \\
&& \sum_{\textit{h}} \upchi_{\textit{h}}(\p)\bar{\upchi}_{\textit{h}}(\p)  =  \frac{\big(|\alpha|^2+|\beta|^2\big)}{2}\gamma_{\mu}p^{\mu}, \label{spinsums2}
\end{eqnarray}
\indent As we are going to see, in order to compute the Hamiltonian and verify its properties, one must consider the non-vanishing projections associated with the Dirac prescription for the dual field\footnote{Note it is also possible to map both notations, and the result above can clearly be written in terms of $\xi(\textbf{p})$ and $\chi(\textbf{p})$, as it follows  
\begin{eqnarray*}
   &&\bar{\xi}_{1}(\textbf{p})\xi_{2}(\textbf{p})=-\frac{im\big(|\alpha|^2+|\beta|^2\big)}{2} = -\bar{\chi}_{1}(\textbf{p})\chi_{2}(\textbf{p}),
    \\
   && \bar{\xi}_{2}(\textbf{p})\xi_{1}(\textbf{p})=\frac{im\big(|\alpha|^2+|\beta|^2\big)}{2} =- \bar{\chi}_{2}(\textbf{p})\chi_{1}(\textbf{p}).
\end{eqnarray*}. However, for clarity in the notations, we will maintain the standard used in equations \eqref{cruzado0} and \eqref{cruzado}.}
\begin{eqnarray}
   &&\bar{\lambda}^{S}_{\{+,-\}}(\textbf{p})\lambda^S_{\{-,+\}}(\textbf{p})=-\frac{im\big(|\alpha|^2+|\beta|^2\big)}{2} = -\bar{\lambda}^{A}_{\{+,-\}}(\textbf{p})\lambda^A_{\{-,+\}}(\textbf{p}),\label{cruzado0}
    \\
   && \bar{\lambda}^{S}_{\{-,+\}}(\textbf{p})\lambda^S_{\{+,-\}}(\textbf{p})=\frac{im\big(|\alpha|^2+|\beta|^2\big)}{2} =- \bar{\lambda}^{A}_{\{-,+\}}(\textbf{p})\lambda^A_{\{+,-\}}(\textbf{p}),
\label{cruzado}\end{eqnarray}
implying that although the norms vanish, there are non-trivial crossed products. These relations concern the components of the new basis associated with Wigner degeneracy.\\
\indent Then, even though in the reference \cite{jcap} it is argued that the entire mass term to be highlighted in \eqref{lagrangiana} vanishes, we must provide a more careful definition of the situation. Despite the fact that the norms of the Elko spinors vanish if the Dirac dual is considered, there are still crossed products that guarantee that the scalar formed by the quantum field and its dual does not equal zero since it is composed of a linear combination of Elko spinors associated with oscillatory operator coefficients.\\
\indent Remarkably, the only uncoupled equation fulfilled by the singular spinors is the Klein-Gordon one
\begin{eqnarray}\label{kg}
   (p^2-m^2)\xi_{\textit{h}}(\p)=0 \quad \mbox{and} \quad (p^2-m^2)\upchi_{\textit{h}}(\p)=0.
\end{eqnarray}
\indent Then, the usual Lagrangian encoding the physics highlighted above reads \cite{mdobook}
\begin{eqnarray}\label{lagrangiana}
 \mathcal{L} = \partial_\mu \bar{\uplambda}(x)\partial^\mu\uplambda(x)-m^2\bar \uplambda(x)\uplambda(x) 
\end{eqnarray}
now in terms of the Dirac instead of the proper dual used for singular spinors.\\
\indent We shall now make a pivotal observation regarding the theory. Once the right-hand side of the spin sums are not proportional, or unitarily connected, to the momentum–space wave operators in equation \eqref{kg}, it will affect the particle interpretation, locality, the propagator structure, and the Hamiltonian properties. Moreover, the fulfillment of the optical theorem \cite{weinberg1}, intimately associated with (pseudo) unitarity, relies upon this correspondence. Since the reference  \cite{jcap} already pointed out that the use of such dual may be possibly problematic for mass dimension one fields, we provide an analysis to show that it indeed is and also explicitly reveal which are these problems. It provides a more solid background for the successful research in the field.

The structures introduced above are in sharp contrast with the standard Dirac case, for particles, 
\begin{eqnarray}
\sum_{\textit{s}=1}^2 \psi^P_{\textit{s}}(\p)\bar{\psi}^P_{\textit{s}}(\p) = \gamma_{\mu}p^{\mu}+m\I    
\end{eqnarray}
and for anti-particles
\begin{eqnarray}
\sum_{\textit{s}=1}^2 \psi^A_{\textit{s}}(\p)\bar{\psi}^A_{\textit{s}}(\p) = \gamma_{\mu}p^{\mu}-m\I,    
\end{eqnarray}
where for this specific case the spin sums determine the wave operator and also determine the
structure of the Feynman–Dyson propagator. Therefore, we see that in the case of Dirac spinors, this dual structure is well-defined, ensuring a consistent and relevant physical interpretation.\\
\indent As a last comment for this section, it is worth mentioning that the proper Elko dual prescription established in \cite{aaca, dharamnpb, jhepgabriel}, defined as $\gdualn{ \chi_h }(\p)=-\big(\mathcal{P}\chi_h(\p)\big)^\dagger\gamma_0$ and $\gdualn{ \xi_h }(\p)=\big(\mathcal{P}\xi_h(\p)\big)^\dagger\gamma_0$, is such that rewriting the Lagrangian \eqref{lagrangiana} in terms of this adjoint implies the following symmetry \cite{aaca} 
\bea  \uplambda(x)\to e^{i\gamma_5}\uplambda(x)\eea
meaning that, just in this specific case, the associated action does not couple the different chiral components of the spinor field. Even though the conservation of the respective associated current may be anomalous at the quantum level, as happens in the standard model, it indicates that mass dimension one fields based on Elko with the correct dual may be a suitable candidate for the so-called sterile neutrino \cite{sterile}. Based on the latter discussion, we are the first researchers to notice that the Lagrangian can be cast in the following form
\begin{eqnarray}
 \mathcal{L} = \partial_\mu \gdualn{\uplambda}_{\mathrm{R}}(x)\partial^\mu\uplambda_{\mathrm{R}}(x)-m^2\gdualn{\uplambda}_{\mathrm{R}}(x)\uplambda_{\mathrm{R}}(x) +\partial_\mu \gdualn{\uplambda}_{\mathrm{L}}(x)\partial^\mu\uplambda_{\mathrm{L}}(x)-m^2\gdualn{\uplambda}_{\mathrm{L}}(x)\uplambda_{\mathrm{L}}(x),
\end{eqnarray}
or in a compact fashion
\begin{eqnarray}
\mathcal{L} = \mathcal{L}_{\mathrm{R}}+\mathcal{L}_{\mathrm{L}},    \end{eqnarray}
in which the chiral components are defined as $ \uplambda_{{\mathrm{R}}/{\mathrm{L}}}(x)\equiv \frac{1}{2}\big( \boldsymbol{1}\pm \gamma_5\big)\uplambda(x)$. This is another deep fundamental difference implicated by the choice of the adjoint prescription. It is interesting to note that the separation between chiral components may be violated if one adds a legitimate renormalizable \footnote{The coupling constant $g$ is dimensionless since all fields in the interaction Lagrangian have mass dimension one nature. } interaction with a scalar mediator like $\mathcal{L}^{\phi \uplambda}=g \phi(x)\gdualn{\uplambda}(x)i\slashed{\partial}\uplambda(x)$, for example, implying in mixing between left and right components; such an important outcome cannot be reached if one employs the Dirac dual structure.

\section{ Problems in the Particle interpretation}\label{3}

\indent Throughout the following subsections, we analyse the main physical consequences of using the Dirac dual structure for singular spinors. First, we calculate the propagator considering the vacuum average of the time-ordered product of the mass dimension one field operator. Then, we obtain the inverse of the theory's differential operator and verify that they differ, implying in a problematic mismatch. In sequence, we also prove that the only Lagrangian formulation that may overcome the mentioned problem is necessarily non-local.\\
\indent Finally, we compute the Hamiltonian operator using the Dirac prescription for the dual obtaining a set of problematic features. They include instabilities and the fact that the one-particle states are no longer eigenstates of the Hamiltonian operator.

\subsection{Mismatch between Propagators and the issue of non-locality }

For now on, we focus just on the case of Elko spinors since this is the most suitable structure to define a dark matter candidate due to its neutral nature.\\
One manner to calculate the Feynman-Dyson propagator is through the time-ordered product of the $\uplambda(x)$ and $\bar{\uplambda}(x)$ operators
\begin{eqnarray}\label{fdpropagator}
iS_{\textrm{FD}}(x-x^{\prime})=\langle 0 \vert \uplambda(x)\bar{\uplambda}(x^{\prime}) \vert 0 \rangle  \theta(x^{0}-x^{\prime 0})- \langle 0 \vert {\uplambda}(x^{\prime})\bar \uplambda(x) \vert 0 \rangle \theta(x^{\prime 0}-x^{0}),
\end{eqnarray}
where the Heaviside function, $\theta(t)$, read
\begin{eqnarray}
\theta(t)\!=\!-\frac{1}{2\pi i}\int_{-\infty}^{\infty} ds\; \frac{e^{-ist}}{s+i\epsilon}.
\end{eqnarray}

Considering the spin sums from \eqref{spinsums1}, \eqref{spinsums2}, and the standard action of the creation and annihilation operators,  the Elko propagator with Dirac dual reads
\begin{align}\label{propagadorfases}
S_{\textrm{FD}}(x^\prime-x)=\int\frac{\text{d}^4 p}{(2 \pi)^4m}\,
e^{-i p_\mu(x^{\prime\mu}-x^\mu)}
\frac{  \gamma_{\mu}p^{\mu}   }{p_\mu p^\mu -m^2 + i\epsilon}.
\end{align}

Now, one can also calculate the propagator taking into account the equations of motion \eqref{kg}. As usual, it is the inverse of the Lagrangian differential operator \eqref{lagrangiana}
Thus, the propagator reads
\begin{eqnarray}\label{prop-classico}
G_{FD}(x-x^{\prime})= \int\frac{d^4p}{(2\pi)^4}e^{-i p_\mu(x^{\prime\mu}-x^\mu)}\frac{1}{p^2-m^2+i\epsilon}.
\end{eqnarray}
We emphasize an important outcome from these results. Note that the calculation of the propagator through the two procedures above provides divergent results, see eq.\eqref{propagadorfases} and eq. \eqref{prop-classico}. For example, obtaining the propagator employing the Schwinger-Dyson equations \cite{mdobook} leads to a different result as compared to the one directly derived through the time-ordered product of the quantum field operators \eqref{campoquanticofinal}. It implies serious obstructions to the consistency of the theory. Namely, this mismatch leads to a violation of the optical theorem, meaning that (generalized) unitarity is not compatible with this specific formulation of a mass dimension one fermionic field.\\
\indent Therefore, the point is that every assertion we made on the field locality depends explicitly on the correct definition of the dual structure in compliance with the recent achievements from \cite{rodolfonogo, rrsingular2024}. The problematic situation concerning the propagators is a direct consequence of using the Dirac adjoint instead of the proper Elko dual.

\subsubsection{Implications in non-locality}
\indent If one insists on reformulating the theory in such a way that both approaches lead to the same propagator, the action should be replaced by \footnote{Whose structure can be proven to be Hermitian.}
\bea   \mathcal{S}=- \int d^4x \bar{\uplambda}(x)\big(\Box+m^2\big)_xi\slashed{\partial}_x\int d^4x'\mathcal{D}(x-x')\lambda(x') \eea
with 
\bea \mathcal{D}(x-x')=\int \frac{d^4p}{(2\pi)^4}e^{-i(x_\mu-x'_\mu).p^\mu} \frac{m}{p^2+i\epsilon}\eea
\indent From variations on the Schwinger-Dyson quantum equations of motion,  one obtains
\bea  \big(\Box+m^2\big)_xi\slashed{\partial}_x\int d^4x'\mathcal{D}(x-x')\langle 0|T\lambda(x')\bar \lambda(y)|0\rangle=\delta^4(x-y)    \eea

implying that the time ordered propagator $\langle 0|T\lambda(x')\bar \lambda(y)|0\rangle$  equals the one calculated in \eqref{propagadorfases}.\\
\indent Although this theory still keeps the solutions in the kernel of the Klein-Gordon differential operator, it is clearly non-local since it is written in terms of the Green function $\mathcal{D}(x-x')$, violating one important quantum field theory foundation.\\

\subsection{On the Hamiltonian instabilities}

Considering the Lagrangian \eqref{lagrangiana}, the canonical momentum operators read $\pi(x)=\dot{\bar {\lambda}}(x)$ and $\bar \pi(x)=\dot{\lambda}(x)$, given operators satisfy the canonical anticommutation relations \cite{jhepgabriel,physrep}, enabling one to compute the Hamiltonian
\bea  H=\int d^3x\Big[\partial_t\bar{\lambda}\partial_t\lambda-\partial^i\bar{\lambda}\partial_i\lambda+m^2\bar{\lambda}\lambda\Big] \eea
\indent Using the relations expressing the crossed products associated with the Dirac dual \eqref{cruzado}, considering the new basis associated with the Wigner degeneracy, and the anti-commutation relations, the normal ordered Hamiltonian operator can be derived
\begin{multline} :H: =i\sum_{\textit{I}, \textit{K}=+,-}\int \frac{d^3p}{(2\pi)^3}E(\p) \varepsilon_{\textit{K} \textit{I}}\left({c^{\dagger S}}_I(\p) c_{K}^S(\p) -c^{\dagger A}_I(\p) c_{K}^A(\p) +d^{\dagger A}_{K}(\p) d_I^A(\p) -d^{\dagger S}_{K}(\p) d_{I}^S(\p) \right),  \end{multline}
with $E(\p)=\sqrt{\p^2+m^2}$, $I=+,-$ and $\varepsilon_{+-}=-\varepsilon_{-+}=1$.\\
\indent The operator is Hermitian. The sum is over the symbols $+,-$ that refers to the first helicity labels of the (anti) self-conjugate spinors composing the new basis \eqref{wig1} and \eqref{wig2}. The ${c_I^\dagger}^{S/A}(\p)$ denotes the creation operator of the particle sector of the new basis associated with spinors classified by the $S$ or $A$ labels regarding self or anti-self conjugate structures, given in \eqref{wig1}-\eqref{wig2}, respectively. An analogous interpretation is valid for the anti-particle sector associated with ${d_I^\dagger}^{A/S}(\p)$. Due to the anti-linearity of the charge conjugation operator, the particle and anti-particle sectors of the new basis, although composed of both $A$ and $S$ spinors, can be arranged in such a way to have a definite respective conjugacy in the Elko phase $\alpha=i$ and $\beta=1$.\\
\indent The standard particle interpretation is no longer valid since the one-particle states
\bea c^\dagger_h(\p)|0\rangle \quad , \ \quad d^\dagger_h(\p)|0\rangle   \eea
with $h=1,.., 4$ are not eigenstates of the Hamiltonian operator.\\
\indent The eigenstates are the following
\begin{multline}{A_I^\dagger}^{A/S}(\p)|0\rangle =\big({c_I^\dagger}^{A/S}(\p)+i\varepsilon_{II^\prime}{c_{I^\prime}^\dagger}^{A/S}(\p)\big)|0\rangle \ ,\quad  {B_I^\dagger}^{A/S}(\p)|0\rangle=\big({{d_I^\dagger}}^{A/S}(\p)+i\varepsilon_{I I^\prime}{d_{I^\prime}^\dagger}^{A/S}(\p) \big)|0\rangle \end{multline}
in which ${A_I^\dagger}^{A}(\p)|0\rangle$ and ${B_I^\dagger}^{A}(\p)|0\rangle$
have negative energy. The remaining ones have positive energy. Then, we demonstrate that considering the Dirac dual for the Elko spinors may lead to problematic situations regarding instabilities, for example.\\
\indent The correct map between the labels considered in this subsection and the components associated with the basis of \eqref{wig1} reads $c^{\dagger S}_{+}(\p)$ $\equiv$ $c^\dagger_{1}(\p)$, $c^{\dagger S}_{-}(\p)$ $\equiv$ $c^\dagger_{2}(\p)$, $c^{\dagger A}_{+}(\p)$ $\equiv$ $c^\dagger_{3}(\p)$, and $c^{\dagger A}_{-}(\p)\equiv c^\dagger_4(\p)$, with an analogous association holding for the anti-particle sector.\\
\indent Additionally, we define which kind of spinor results from the contraction of the Hamiltonian eigenstate and the quantum field operator
\indent  \[
\wick{
       \c2 \uplambda(x)
       \c2 {A^{\dagger}_{\pm}}^{S/A}(\p) \vert
          0
  }
\rangle=e^{-ip_\mu x^\mu}\Lambda_{\pm}^{S/A}(\p)\vert 0\rangle
\]
with $\Lambda_{\pm}^{S}(\p)=\uplambda^{S}_{\{\pm,\mp\}}(\p)\pm i\uplambda^{S}_{\{\mp,\pm\}}(\p)      $ and $\Lambda_{\pm}^{A}(\p)=-i\big(\uplambda^{A}_{\{\pm,\mp\}}(\p)\pm i\uplambda^{A}_{\{\mp,\pm\}}(\p)\big)      $. \\
\indent The resulting spinor field associated with the eigenstates obey the following equations
\bea  \slashed{p}\Lambda^S_\pm(\p)=m\Lambda^S_\pm(\p) \ , \qquad \slashed{p}\Lambda^A_\pm(\p)=-m\Lambda^A_\pm(\p)  \eea
meaning that they defined eigenstates of the parity operator, in contrast with the Elko spinor definition. This is to be expected considering the structure of such eigenstates as well as the fact that the Lounesto classes are not invariant under the algebraic sum of spinors within a given specific class \cite{lounestolivro}. In this manner, one also verifies that the neutral nature is no longer valid since 
\bea  \gamma_2{\Lambda^S_+}^*(\p)=-i\Lambda^S_-(\p)     \eea
the action of the charge conjugation operator in the particular example above ladders between the $\Lambda^S_I(\p)$ spinors. All the discussed features reinforce the idea that the Dirac adjoint prescription is not suitable for spinors that are not eigenstates of the generalized parity operator $\mathcal{P}$ \cite{speranca}.  

\indent It is also interesting to consider an alternative route for this discussion. Since the canonical analysis reveals that the momenta are proportional to the time derivatives of the fields and their duals, the correspondence principle \cite{nakahashi} implies the relation below
\bea \Big\{\uplambda(\vec x,t),\bar{\uplambda}(\vec y,t)\Big\}=0\eea
for the quantum field. However, the standard anti-commutator algebra \eqref{anticomutador} cannot fulfill this condition. In fact, it leads to 
\bea \Big\{\lambda(\vec x,t),\bar{\lambda}(\vec y,t)\Big\}=\frac{\gamma_0}{m}\delta^3(\vec x-\vec y) \eea
being clearly incompatible with the correspondence principle.\\
\indent Interestingly, this quantization step can be successfully carried out if one replaces the non-vanishing part of the anti-commutator algebra by  
\begin{equation}
\Big\{{c_I}^{S/A}(\p),c^{\dagger S/A}_{I^{\prime}}(\p^{\prime})\Big\} =\pm i \varepsilon_{II^{\prime}}(2\pi)^3 \delta^3(\p-\p^\prime), \quad \Big\{{d_I}^{A/S}(\p),d^{\dagger A/S}_{I^{\prime}}(\p^{\prime})\Big\} =\pm i \varepsilon_{II^{\prime}} (2\pi)^3 \delta^3(\p-\p^\prime)  \end{equation}
corresponding to the following Hamiltonian \footnote{Which still has a set of negative eigenvalues.}
\begin{multline} :H: =\sum_{\textit{I}=+,-}\int \frac{d^3p}{(2\pi)^3}E(\p) \left(c^{\dagger S}_{I}(\p) c_{I}^S(\p) +c^{\dagger A}_{I}(\p) c_{I}^A(\p) +d^{\dagger A}_{I}(\p) d_I^A(\p) +d^{\dagger S}_{I}(\p) d_{I}^S(\p) \right),  \end{multline}
evincing that the formulation is necessarily incorrect, such a feature is a direct consequence of the dual employed along the calculations. Despite these efforts, considering the redefined commutator algebra one concludes that the one-particle states are still not eigenstates of the Hamiltonian operator. This operator also has negative eigenvalues.  Moreover, the norm of the one-particle states vanishes according to this quantization prescription. 

We conclude this section by emphasizing that, as we have seen, the dual is a fundamental structure for any spinorial theory. Throughout this work, we have highlighted numerous flaws and severely problematic consequences that arise if such a structure is not correctly defined. One final relevant point that should be mentioned, and which is fully in line with all the results presented here, is that the only way to ensure Hermiticity \cite{elkoherm} and a way to connect gravity and supergravity with a generalized Elko theory \cite{supergrav} are correctly developed by using the appropriated dual structure.

\section{Concluding remarks}\label{4}
\indent Throughout these sections, we provided several different careful discussions on the fact that the Dirac adjoint does not support a well-defined field quantization based on singular spinor expansion coefficients. It implies the necessity of a generalized dual prescription to describe the physics associated with such mathematical objects commonly addressed as dark matter candidates. Therefore, the paper provided an additional contribution to establishing the pertinence of the historical route followed in the dark spinor research. As a future perspective, we can search for the set of symmetries allowed for each kind of adjoint formulation and highlight the specific contents of the generalized Elko dual. Besides that, we can also investigate the properties of adjoint formulation of Elko in the context of supergravity in which an extra field enters in the game \cite{supergrav}

\section{Acknowledgements}
GBdG thanks the S\~ao Paulo Research Foundation FAPESP (2021/12126-5) for the financial support and also the hospitality offered by UFTM. RJBR thanks the generous hospitality offered by UNIFAAT and Professor Renato Medina. LF thanks the Next Generation EU project ``Geometrical and Topological effects on Quantum Matter (GeTOnQuaM)'' for the financial support.


\begin{thebibliography}{40}

\bibitem{jcap}D.~V.~Ahluwalia and D.~Grumiller,
Spin half fermions with mass dimension one: Theory, phenomenology, and dark matter,
JCAP \textbf{07} (2005) 012
[arXiv:hep-th/0412080 [hep-th]].

\bibitem{dharamprd}
D.~V.~Ahluwalia and D.~Grumiller, Dark matter: A spin one half fermion field with mass dimension one?, Phys.Rev.{\bf D72}, 067701 (2005).

\bibitem{polarform2020} L. Fabbri and R. J. B. Rogerio, Polar form of spinor fields from regular to singular: the flag-dipoles, The European Physical
Journal C {\bf 80}, 8 (2020).

\bibitem{elkopolar} L. Fabbri, ELKO in polar form, The European Physical Journal Special Topics 229, 2117 (2020).

\bibitem{chenggeneral} C.-Y. Lee, Spin-half mass dimension one fermions and their higher-spin generalizations, The European Physical Journal
Special Topics 229, 2003 (2020).

\bibitem{chenglagrangian} C.-Y. Lee, A lagrangian for mass dimension one fermionic dark matter, Physics Letters B 760, 164 (2016).

\bibitem{juliodoublets}
F. A. da Silva Barbosa, J. M. Hoff da Silva, Non-standard Wigner doublets, EPL {\bf 144}, 5, 54001 (2023).

\bibitem{saulo1} S. H. Pereira, Degeneracy pressure of mass dimension one fermionic fields and the dark matter halo of galaxies, International
Journal of Modern Physics D 31 (2022).

\bibitem{saulo2} S. H. Pereira, M. E. S. Alves, and T. M. Guimar˜aes, An unified cosmological evolution driven by a mass dimension one
fermionic field, The European Physical Journal C 79 (2019).

\bibitem{saulo3} S. H. Pereira and R. S. Costa, Partition function for a mass dimension one fermionic field and the dark matter halo of
galaxies, Modern Physics Letters A 34, 1950126 (2019).

\bibitem{cylcosmelkology}
Cheng-Yang Lee, Haomin Rao, Wenqi Yu, Siyi Zhou, Cosmelkology: Elko fermions in FLRW space-time, 	arXiv:2402.05623 [hep-th] (2024).


\bibitem{roldao2023} G. de Gracia, A. Nogueira, and R. da Rocha, Fermionic dark matter-photon quantum interaction: A mechanism for
darkness, Nuclear Physics B 992, 116227 (2023).

\bibitem{roldaofermionic} G. B. de Gracia and R. da Rocha, Fermionic dark matter interaction with the photon and the proton in the quantum
field-theoretical approach and generalizations (2022).

\bibitem{julioperturbative} W. Carvalho, M. Dias, A. C. Lehum, and J. M. H. da Silva, Perturbative aspects of mass dimension one fermions nonminimally
coupled to electromagnetic field (2023).

\bibitem{chengyukawa} M. Dias and C.-Y. Lee, Constraints on mass dimension one fermionic dark matter from the Yukawa interaction, Physical
Review D 94 (2016).

\bibitem{wigner1} Wigner E P, Unitary representations of the inhomogeneous Lorentz group including reflections, 1964
Group Theoretical Concepts and Methods in Elementary Particle Physics (Lectures of the Istanbul
Summer School of Theoretical Physics (1962)) ed F G¨ursey (New York: Gordon and Breach) 

\bibitem{wigner2} Wigner E P, On unitary representations of the inhomogeneous Lorentz group, 1939 Ann. Math. 40 149
Wigner E P, 1989 Nucl. Phys. Proc. Suppl. 6 9 (Reprinted)

\bibitem{dharamnpb} D. V. Ahluwalia, J. M. Hoff da Silva, C.-Y. Lee, Mass dimension one fields with Wigner degeneracy: A theory of dark matter, Nuclear Physics B, {\bf 987}, 116092 (2023).

\bibitem{weinberg1} S. Weinberg, {\it The Quantum Theory of Fields}, Vol. I: Foundations, Cambridge University Press, New York, (1996).

\bibitem{ryder}
L. H. Ryder, \textit{Quantum Field Theory}, Second Edition, Editorial Cambridge University Press, New York (1996).

\bibitem{pamdirac}
P. A. M. Dirac Paul, A theory of electrons and protons, Proc. R. Soc. Lond. A, 126360–365 (1930).

\bibitem{rrdirac1}
C. H. Coronado Villalobos, R. J. Bueno Rogerio, The connection between Dirac dynamic and parity symmetry, Europhysics Letters {\bf 116}, 60007 (2016).

\bibitem{rrdirac2}
 R. J. Bueno Rogerio, C. H. Coronado Villalobos, Non-standard Dirac adjoint spinor: The emergence of a new dual, Europhysics Letters {\bf 121}, 21001 (2018).

 \bibitem{dharamnogo}
D. V. Ahluwalia, Evading Weinberg's no-go theorem to construct mass dimension one fermions: Constructing darkness, Europhys. Lett. {\bf 118}, 6, 60001 (2017).

 \bibitem{rodolfonogo}
 R. J. Bueno Rogerio, J. M. Hoff da Silva, C. H. Coronado Villalobos, Regular spinors and fermionic fields, Physics Letters A {\bf 402}, 127368 (2021).


\bibitem{dharambosons} D. V. Ahluwalia, Spin-half bosons with mass dimension three half: towards a resolution of the cosmological constant problem, EPL {\bf 131}, 41001 (2020).

\bibitem{dharamspinstatistic} D. V. Ahluwalia and C.-Y. Lee, Spin-half bosons with mass dimension three-half: Evading the spin-statistics theorem, EPL {\bf 140}, 24001 (2022).

\bibitem{ramond}
P. Ramond, \textit{Field Theory: A modern primer}, Second Edition, Editorial Addison Wesley-Publishing, California (1989).

\bibitem{rodolfoconstraints}
R. J. Bueno Rogerio, Constraints on mapping the Lounesto’s classes, Eur. Phys. J. C {\bf 79}, 929 (2019).
 
\bibitem{rjdual} J. M Hoff da Silva, R. J. Bueno Rogerio and N. C. R. Quinquiolo. Spinorial discrete symmetries and adjoint structures. Phys. Let. A, {\bf 452}, 128470 (2022).

\bibitem{rrtaka} R. J. Bueno Rogerio, R. T. Cavalcanti, J. M. Hoff da Silva, et.al.,
Revisiting Takahashi's inversion theorem in discrete symmetry-based dual frameworks, 	Physics Letters A, {\bf 481}, 129028 (2023).

\bibitem{propagatormpla} R. J. Bueno Rogerio, Singular spinors and their connection, Mod. Phys. Let. A, {\bf 36}, 2150093 (2021).

\bibitem{propagatorbsm} R. J. Bueno Rogerio, L. Fabbri, Propagators Beyond The Standard Model, 	Adv. Appl. Clifford Algebras {\bf 33}, 39 (2023).

\bibitem{rjtau} R. J Bueno Rogerio and J. M. Hoff da Silva, The local vicinity of spin sums for certain mass-dimension-one spinors. EPL, {\bf 118}, 10003 (2017).

\bibitem{chengtipo4} Lee, C-Y. Fermionic degeneracy and non-local contributions in flag-dipole spinors and mass dimension one fermions. Eur. Phys. Jour. C, {\bf 81}, 1 (2021).

\bibitem{out} 
J. M. Hoff da Silva, R. T. Cavalcanti, Revealing how different spinors can be: the Lounesto 
spinor classification, \textit{Mod.Phys.Lett.A}\textbf{32}, 1730032 (2017).
 
\bibitem{jrold1}
J .M. Hoff da Silva, R. da Rocha,
Unfolding Physics from the Algebraic Classification of Spinor Fields, 
\textit{Phys. Lett. B}\textbf{718}, 1519 (2013).

\bibitem{rog1} J. M. Hoff da Silva and R. T. Cavalcanti, Further investigation of mass dimension one fermionic duals, Phys. Lett. A {\bf 383}, 1683 (2019).

\bibitem{rog2}
R.T.Cavalcanti, J. M. Hoff da Silva, Unveiling Mapping Structures of Spinor Duals, 
\textit{Eur.Phys.J.C}\textbf{80}, 325 (2020).

\bibitem{beyondlounesto}
C. H. Coronado Villalobos, R. J. Bueno Rogerio, A. R. Aguirre, et al., On the generalized spinor classification: beyond the Lounesto’s classification, Eur. Phys. J. C {\bf 80}, 228 (2020). 

\bibitem{aaca} D. V. Ahluwalia, {\it The theory of local mass dimension one fermions of spin one half}, Adv. Appl. Clifford Algebras {\bf 27}, 2247 (2017).

\bibitem{speranca}
L. D. Sperança, \emph{Int. J. Mod. Phys.} \textbf{D 2}, 1444003 (2014).

\bibitem{lounestolivro}
P. Lounesto, \textit{Clifford algebras and spinors}, Second Edition, Editorial Cambridge University Press, Cambridge (2002).

\bibitem{mdobook}
D. V. Ahluwalia, {\it Mass Dimension One Fermions} (Cambridge Monographs on Mathematical Physics). Cambridge University Press, Cambridge, (2019).

\bibitem{elkoherm}
G. B. de Gracia, et.al. On Wigner Degeneracy in Elko theory: Hermiticity and Dark Matter Phenomenological Constraints, paper in progress (2024). 

\bibitem{supergrav}
J.A. Nieto and E. A. León, Towards a gravitational and supergravitational Elko theory, Int. J. Mod.  Phys.  A {\bf 37}  06, (2022). 

\bibitem{nieto1}
J. A. Nieto, Generalized Elko Theory, Mod. Phys. Let. A {\bf  34}, No. 26, 1950211 (2019).

\bibitem{nieto2}
J. A. Nieto, Higher Dimensional Elko Theory, Rev. Mex. Fis. {\bf 60}, 371-375 (2014).

\bibitem{elkostates} J. M. Hoff da Silva, and R. J. Bueno Rogerio, {\it Massive spin-one-half one-particle states for the mass-dimension-one fermions.} EPL {\bf 128}, 11002 (2019).


\bibitem{rcampla} 
C. H. Coronado Villalobos, R. J. Bueno Rogerio, A. R. Aguirre, Spinorial structures, discrete symmetries and some consequences, Mod. Phys. Lett. A {\bf 36}, 18, 2150129 (2021).

\bibitem{jhepgabriel}
D. V. Ahluwalia, et.al. Irreducible representations of the inhomogeneous Lorentz group with two-fold Wigner degeneracy,  J. High Energ. Phys. {\bf 75} (2024).


\bibitem{rrsingular2024}
R. J. Bueno Rogerio, C. H. Coronado Villalobos, Singular spinors as expansion coefficients of local spin-half fermionic and bosonic fields: On the two-fold Wigner degeneracy, Physics Letters A {\bf 498}, 129348 (2024).

\bibitem{sterile} B. Dasgupta and J. Kopp, Sterile Neutrinos,  Phys. Rept. {\bf 928}, 1-63 (2021).

 \bibitem{nakahashi}
N. Nakanishi, and I. Ojima, \emph{Covariant operator formalism of gauge theories and quantum gravity}, World Scientific Publishing Co. Pte. Ltd., Editorial  Utopia Press, Singapore (1990).

\bibitem{physrep}
D. V. Ahluwalia, et.al. Mass dimension one fermions: Constructing darkness, Phys. Rept. {\bf 967}, 1-43 (2022). 




\end{thebibliography}
\end{document}